\documentclass[fleqn,usenatbib]{mnras}
\usepackage{graphicx,aas_macros,bm,amssymb,amsmath,times}
\usepackage{xfrac}
\numberwithin{equation}{section}
\usepackage{xcolor}
\usepackage{gensymb}

\allowdisplaybreaks

\title[GRSPH simulations of TDEs]{Disc formation from tidal disruption of stars on eccentric orbits by Kerr black holes using GRSPH}

\author[Liptai et al.]{
David Liptai,$^{1}$\thanks{E-mail: david.liptai@monash.edu}
Daniel J. Price,$^{1}$\thanks{E-mail: daniel.price@monash.edu}
Ilya Mandel,$^{1,2,3}$
Giuseppe Lodato$^{4}$
\\
$^1$Monash Centre for Astrophysics (MoCA) and School of Physics and Astronomy, Monash University, Clayton, Vic. 3800, Australia \\
$^2$ The ARC Centre of Excellence for Gravitational Wave Discovery -- OzGrav\\
$^3$ Birmingham Institute for Gravitational Wave Astronomy and School of Physics and Astronomy,\\ University of Birmingham,  B15 2TT, Birmingham, UK\\
$^4$Dipartimento di Fisica, Universit\`a Degli Studi di Milano, Via Celoria 16, Milano 20133, Italy\\
}

\pubyear{2019}

\begin{document}
\label{firstpage}
\pagerange{\pageref{firstpage}--\pageref{lastpage}}
\maketitle

\begin{abstract}
We perform 3D general relativistic smoothed particle hydrodynamics (GRSPH) simulations of tidal disruption events involving 1 $M_\odot$ stars and $10^6 M_\odot$ rotating supermassive black holes.
We consider stars on initially elliptical orbits both in, and inclined to, the black hole equatorial plane.
We confirm that stream-stream collisions caused by relativistic apsidal precession rapidly circularise the disrupted material into a disc.
For inclined trajectories we find that nodal precession induced by the black hole spin (i.e. Lense-Thirring precession) inhibits stream-stream collisions only in the first orbit, merely causing a short delay in forming a disc, which is inclined to the black hole equatorial plane.
We also investigate the effect of radiative cooling on the remnant disc structure.
We find that with no cooling a thick, extended, slowly precessing torus is formed, with a radial extent of 5 au (for orbits with a high penetration factor).
Radiatively efficient cooling produces a narrow, rapidly precessing ring close to pericentre.
We plot the energy dissipation rate, which tracks the pancake shock, stream-stream collisions and viscosity. We compare this to the effective luminosity due to accretion onto the black hole. We find energy dissipation rates of $\sim10^{45}$ erg s$^{-1}$ for stars disrupted at the tidal radius, and up to $\sim10^{47}$ erg s$^{-1}$ for deep encounters.
\end{abstract}


\section{Introduction} \label{sec:intro}
The tidal disruption of stars by supermassive black holes is thought to power bright electromagnetic flares \citep{lidskiiozernoi79,rees88} that are observed at the centre of galaxies \citep[see ][for a review]{komossa15}. This occurs when a star gets too close to the black hole, such that the tidal forces of the black hole exceed the self gravity of the star \citep{hills75}. The subsequent accretion of stellar material onto the black hole proves useful as a tool for probing quiescent supermassive black holes (SMBHs).

The luminosity for such tidal disruption events (TDEs) decays loosely as a power-law $\propto t^{-5/3}$ \citep[e.g.][]{gezaribasamartin08,gezarichornockrest12,auchettlguillochonramirez-ruiz17}, which is consistent with the predicted mass return rate of stellar debris at late times \citep{rees88,phinney89}. After the initial disruption, the star --- initially on a parabolic orbit --- is torn into a long stream, with roughly half the mass being gravitationally bound to the black hole and returning on highly eccentric orbits. In order for the light curve to trace the fallback rate, the stellar debris must circularise, form an accretion disc, and accrete on a timescale much shorter than the return time of the most tightly bound debris \citep{stonekesdencheng19} --- assuming that the observed flare is powered by accretion onto the black hole and not by fallback shocks and circularisation \citep[e.g.][]{lodato12,piransvirskikrolik15,jiangguillochonloeb16}. The manner in which material circularises is still uncertain, but is thought to be due to shock dissipation from the self-intersection of the debris stream, caused by relativistic apsidal precession \citep{rees88,cannizzoleegoodman90,hayasakistoneloeb13,bonnerotrossilodato16}. However, the process of circularisation is further complicated by frame dragging. This causes precession of the orbital plane with each pericentre passage, which can prevent stream self-intersection \citep{guillochonramirez-ruiz15}.

Investigating TDEs numerically presents a significant challenge, since the process involves a large range of length and time scales that need to be resolved from disruption to accretion. In the case of a $1M_\odot$ star on a parabolic orbit disrupted by a $10^6M_\odot$ SMBH, the period of the most bound debris is $\sim\,$$10^3$ times greater than the dynamical time of the star, while the semi-major axis of the most bound debris is $\sim\,$$10^4$ times greater than the stellar radius \citep{stonekesdencheng19}. As a consequence, most previous numerical studies have focused on the details of the initial disruption process, and have not simulated the subsequent debris fallback and predicted disc formation \citep[e.g.][]{evanskochanek89,lagunamillerzurek93,khokhlovnovikovpethick93,frolovkhokhlovnovikov94,dienerfrolovkhokhlov97,lodatokingpringle09,guillochonramirez-ruizrosswog09,guillochonramirez-ruiz13,tejedagaftonrosswog17,gaftonrosswog19}. The relativistic SPH calculations of \citet{ayalliviopiran00} did follow fallback, however they were unable to resolve an accretion disc due to the small number of SPH particles employed. \cite{guillochonmanukianramirez-ruiz14} also followed debris fallback in their simulations, however were only able to form highly elliptic accretion flows since they had no treatment of relativistic effects.
Similarly, \citet{rosswogramirez-ruizhix08} and \citet{shiokawakrolikcheng15} showed little debris circularisation, although they restricted their calculations to a more numerically favourable choice of parameters, namely
the disruption of white dwarfs by intermediate mass black holes.

\citet{hayasakistoneloeb13} were the first to convincingly demonstrate disc formation and debris stream self-intersection due to apsidal precession around SMBHs, using a pseudo-Newtonian gravitational potential in SPH calculations. Similar studies by \citet{bonnerotrossilodato16} and \citet{sadowskitejedagafton16} confirmed their results, although all three studies only considered non-rotating black holes, and stars on initially bound, highly eccentric orbits.
The role that black hole spin plays in circularisation has only recently been investigated numerically by \citet{hayasakistoneloeb16}. Using SPH with a post-Newtonian approximation for the Kerr metric, they showed that nodal precession for highly elliptical orbits out of the black hole spin plane can delay debris circularisation.

In this paper, we investigate debris circularisation and subsequent disc formation around rotating black holes, with particular interest in initial orbits that are out of the black hole spin plane. We use general relativistic smoothed particle hydrodynamics (GRSPH) in a fixed background Kerr metric, allowing us to treat both apsidal precession and Lense-Thirring precession correctly in the strong field limit. As in the studies listed above, we limit ourselves to stars on initially bound, highly eccentric orbits.

This paper is organised as follows. In Section~\ref{sec:method} we describe the method and initial conditions. In Section~\ref{sec:results} we present our results, investigating the effect of black hole spin and inclination, penetration factor and efficiency of radiative cooling. We discuss our results in Section~\ref{sec:discussion} and present our conclusions in Section~\ref{sec:conclusion}.

\section{Method} \label{sec:method}
We simulated the tidal disruption of solar type stars in 3D using GRSPH. We used the general relativistic implementation of \textsc{Phantom} \citep{pricewurstertricco18} as described in \citet{liptaiprice19} to perform our calculations. The calculation is performed in the Kerr metric in Boyer-Lindquist coordinates (with black hole spin vector along the $z$-axis), which describes the spacetime around a rotating black hole for a distant observer, along with Newtonian self-gravity in order to hold the star together during the pre-disruption phase. We set an accretion radius at $r_\mathrm{acc}=5GM_\mathrm{BH}/c^2$ i.e. within the innermost stable circular orbit (ISCO) for a Schwarzschild black hole. Particles that fall within this radius are removed from the simulation. This prevents particles requiring very small time steps close to the horizon, which would significantly hinder code performance.

\begin{table}
\centering
\caption{Parameters for each simulation}
\label{tab:sims}
\begin{tabular}{cccccc}
\hline
No. & $e$  & $\beta$ & $a$   & $\theta$    & cooling \\
\hline
1   & 0.95 & 5       & 0     & $0\degree$  & adiabatic      \\
2  & 0.95 & 5       & 0.99  & $0\degree$  & adiabatic      \\
3  & 0.95 & 5       & 0.99  & $30\degree$ & adiabatic      \\
4  & 0.95 & 5       & 0.99  & $60\degree$ & adiabatic      \\
5  & 0.95 & 5       & 0.99  & $89\degree$ & adiabatic     \\
6  & 0.95 & 1       & 0.99  & $60\degree$ & adiabatic      \\
7   & 0.95 & 5       & 0     & $0\degree$  & isentropic     \\
8  & 0.95 & 5       & 0.99  & $60\degree$ & isentropic     \\
\hline
\end{tabular}
\end{table} 

\subsection{Setup}
We consider the disruption of stars with mass $M_*=1M_\odot$ and radius $R_*=1R_\odot$ by a supermassive black hole with mass $M_\mathrm{BH}=10^6M_\odot$. Each star is placed on an eccentric orbit at apocenter $R_\mathrm{a} = R_\mathrm{p} (1+e)/(1-e)$, with velocity $v_\mathrm{a} = \sqrt{G M_\mathrm{BH}(1-e)/R_a}$, where $e$ is the eccentricity and $R_\mathrm{p}$ is the Newtonian pericentre distance. In practice, the point of closest approach can differ from $R_\mathrm{p}$ by up to $\sim\,$$2 GM_\mathrm{BH}/c^2$ due to relativistic effects, and thus the effective penetration factor $\beta$ can differ by up to $\sim\,$1 for our most penetrating orbits.

We vary the orbital inclination $\theta$, the black hole spin $a$, and the penetration factor $\beta \equiv R_\mathrm{t}/R_\mathrm{p}$, where $R_\mathrm{t}=R_*(M_\mathrm{BH}/M_*)^{1/3}$ is the tidal radius. For our choice of parameters we have
\begin{equation}
   R_\mathrm{p} = 47.1 R_g \left(\frac{\beta}{1}\right)^{-1} \left(\frac{R_*}{R_\odot}\right) \left(\frac{M_\mathrm{BH}}{10^6 M_\odot}\right)^{-2/3} \left(\frac{M_*}{M_\odot}\right)^{-1/3},
\end{equation}
where $R_g\equiv GM_\mathrm{BH}/c^2$.
For orbits in the black hole equatorial plane, we place the star on the $x$-axis, while for inclined trajectories we rotate the plane of orbit up by an angle $\theta$ from the $x$-$y$ plane.

We model the star as an initially polytropic sphere with polytropic index $\gamma=5/3$, using approximately $524$K particles ($\sim\,$$4 \pi 50^3/3$). We initially place the particles on a cubic lattice, and then stretch their radial positions to achieve the desired density profile, as given by the corresponding solution to the Lane-Emden equation, using the `star' setup in \textsc{Phantom} \citep{pricewurstertricco18}. 

We perform our simulations in geometric units where $G=c=M_\mathrm{BH}=1$ but present results in physical units assuming a $10^6M_\odot$ black hole. The corresponding transformations from code to physical units are given by
\begin{align}
   t_{\rm phys} &= 4.926 \mathrm{s} \times t_{\rm code} \left(\frac{M_\mathrm{BH}}{10^6 M_\odot}\right), \\
   R_\mathrm{phys} &= 0.0098 \mathrm{au} \times R_{\rm code} \left(\frac{M_\mathrm{BH}}{10^6 M_\odot}\right).
\end{align}

Table~\ref{tab:sims} shows a list of our simulations presented in this paper and the parameters used for each.
We initially considered a wider set of parameters, including more bound orbits similar to those computed by \citet{bonnerotrossilodato16}, but since most tidal disruptions are expected to involve stars on marginally bound orbits, we omit these calculations from our analysis in the interest of brevity.

\begin{figure*}
   \begin{center}
      \includegraphics[width=\textwidth]{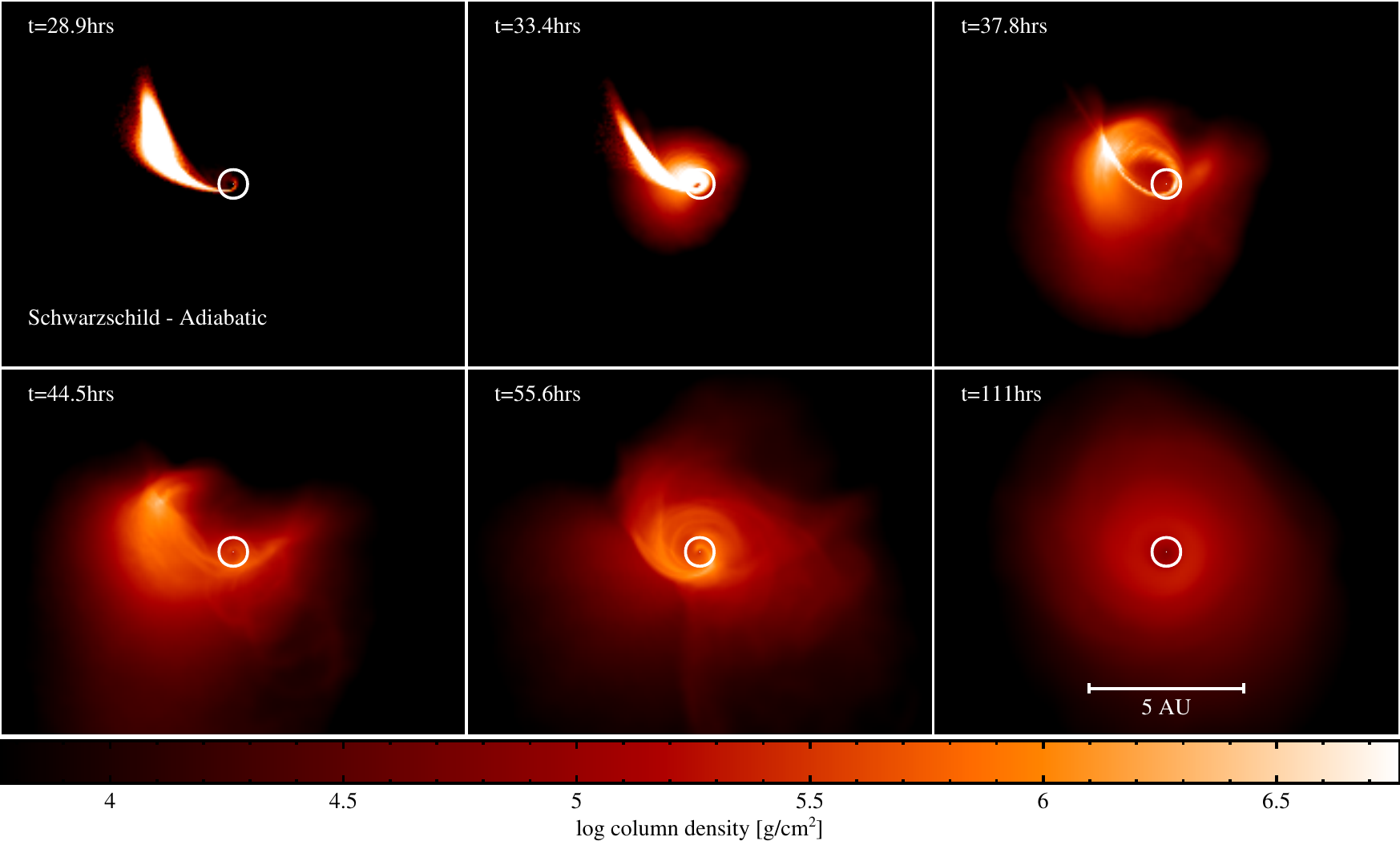}
      \caption{Snapshots of the debris fallback projected in the $x$-$y$ plane for the $e=0.95$, $\beta=5$, $\theta=0\degree$, $a=0$ adiabatic simulation (Sim. 1). Starting at apocentre on the $+x$-axis, the star travels counter-clockwise coming within $<10 R_g$ of the black hole at pericentre (not shown). Apsidal precession advances the orbit by $>90\degree$ (panel 1), which on the second pericentre passage (panel 2) causes the stream to self-intersect and dissipate orbital energy through shocks (panel 3). Over the next few orbital periods, the debris circularises and forms a thick torus (panel 6). White circle indicates the tidal radius $R_\mathrm{t}$.}
      \label{fig:main-schwarzschild}
   \end{center}
\end{figure*}

\begin{figure*}
   \begin{center}
      \includegraphics[width=\textwidth]{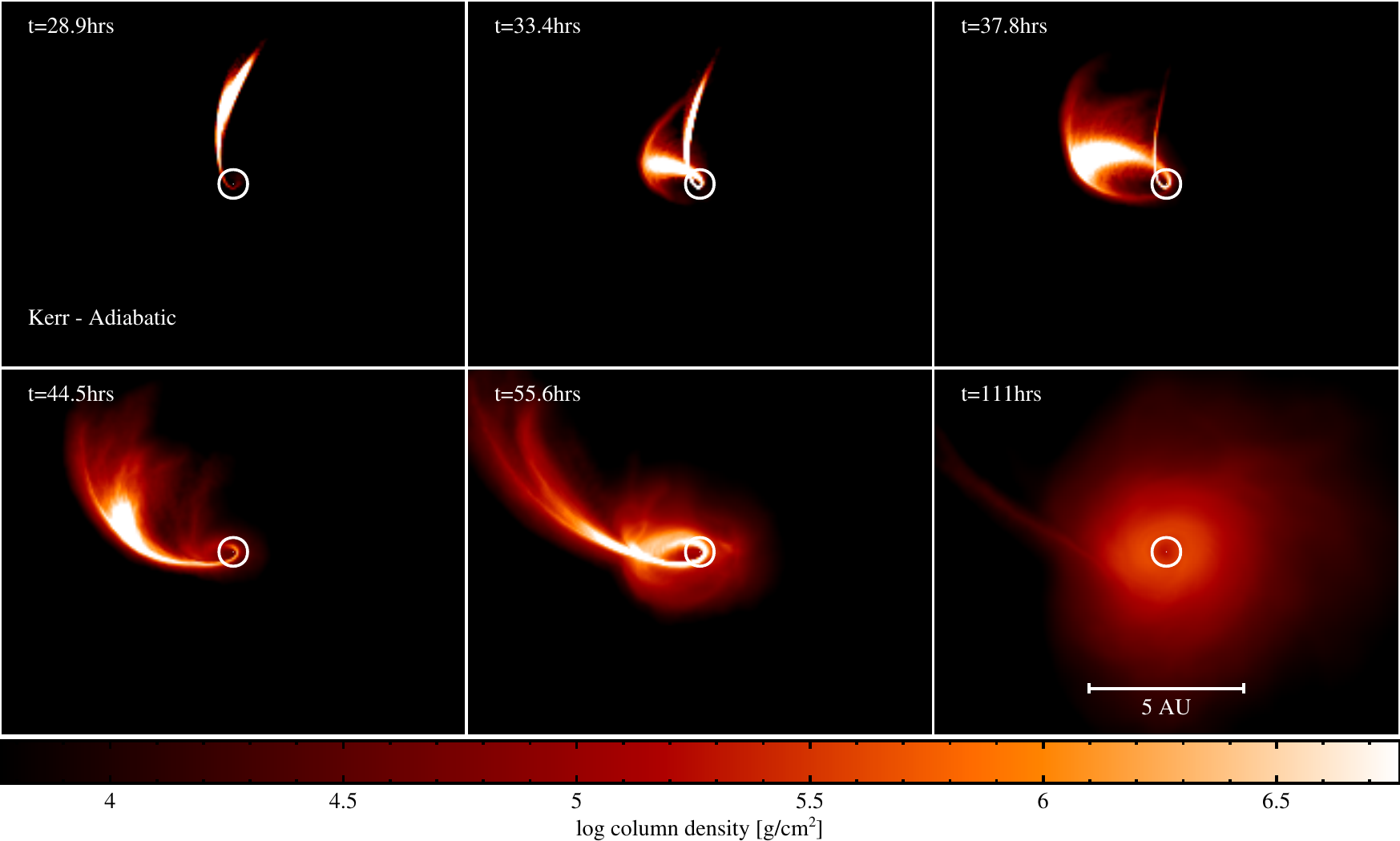}
      \caption{Snapshots of the debris fallback for the $e=0.95$, $\beta=5$, $\theta=60\degree$, $a=0.99$ adiabatic simulation (Sim. 4), projected in the initial orbital plane. Since the black hole is rotating rapidly and the initial orbit does not lie in the spin plane, nodal precession moves the star into a new orbital plane after the first pericentre passage, preventing stream self-intersection early on. However as the stream thickens, a significant portion of the stream self-intersects and a torus is formed. White circle indicates the tidal radius $R_\mathrm{t}$.}
      \label{fig:main-kerr}
   \end{center}
\end{figure*}

\subsection{Accretion luminosity and heating rate} \label{sec:lightcurves}
We derive an effective luminosity from the mass accretion rate \citep[cf.][]{vigneronlodatoguidarelli18}
\begin{equation}
   L_\mathrm{acc} = \epsilon \dot{M} c^2,
\end{equation}
where $\dot{M}$ is the mass accretion rate crossing our inner boundary, and $\epsilon=0.2$ is the efficiency based on our inner accretion radius. 

Our second estimate of the energy release comes from the irreversible dissipation of energy produced by shocks and viscous heating. This is fairly straightforward to compute in our GRSPH code since we evolve an entropy variable $K\equiv P/\rho^\gamma$ according to
\begin{align}
   \frac{d K}{d t} = \frac{K}{u} \left( \frac{du}{dt} - \frac{P}{\rho^2} \frac{d\rho}{dt} \right),
\end{align}
which for an individual particle, denoted by the subscript $a$, involves only the SPH shock heating term $\Lambda^\mathrm{shock}$ \citep[cf. equation 69 of][]{liptaiprice19}
\begin{align} \label{eq:entropy}
   \frac{d K_a}{d t} = \frac{K_a}{u_a} U^0_a \Lambda_a^\mathrm{shock}.
\end{align}
Here $U^0$ is the temporal component of the four velocity, and $\rho$, $P$ and $u$ are the density, pressure and specific thermal energy of the gas, respectively.
Thus we identify the $U^0_a \Lambda_a^\mathrm{shock}$ term as the specific energy generation rate per particle, and hence the total injected luminosity at a given time is
\begin{equation}
   L_\mathrm{shock} = \sum_a m_a U^0_a \Lambda_a^\mathrm{shock},
\end{equation}
where $m_a$ is the particle mass. We emphasise that this does not necessarily reflect the true luminosity evolution since energy would be radiated through a photosphere in practice (see our caveats in Sec.~\ref{sec:discussion}).

\subsection{Radiative cooling} \label{sec:cooling}
In our standard adiabatic calculations the heat produced by shocks (and viscosity) does not escape from the gas i.e. radiative shock cooling is \textit{inefficient}.
We also perform isentropic calculations in Section~\ref{sec:isentropic}, where the shock heating terms are discarded from the entropy evolution in Eq.~\ref{eq:entropy} i.e. radiative shock cooling is \textit{efficient}.
We note that in reality the gas behaviour is somewhere between these two extremes, however without doing careful radiative transfer calculations, it is not obvious how the heat generated from shocks is actually reprocessed and radiated.

\begin{figure*}
   \begin{center}
      \includegraphics[width=\columnwidth]{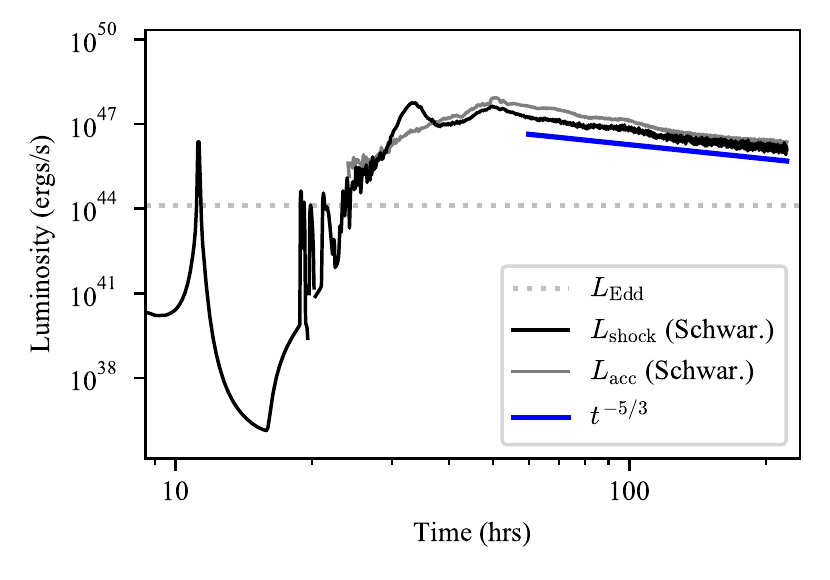}
      \includegraphics[width=\columnwidth]{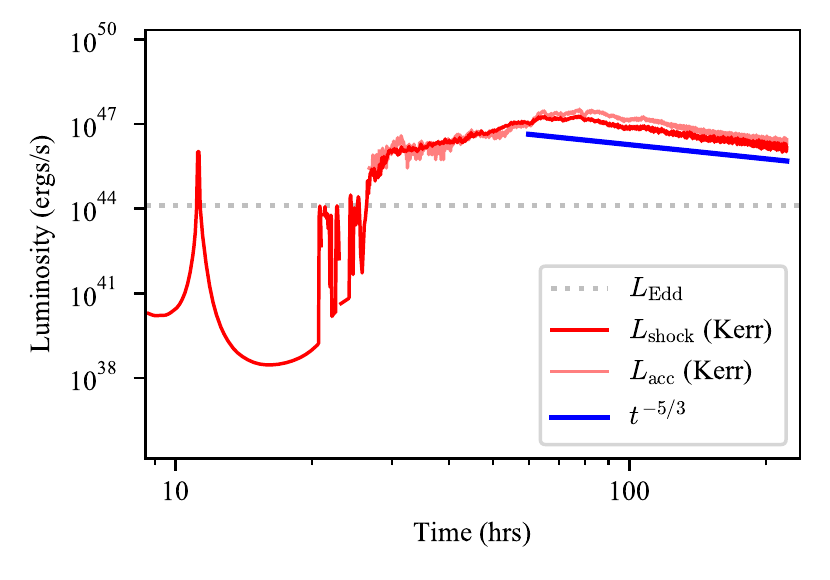}
      \caption{Energy injection rate ($L_\mathrm{shock}$) and accretion luminosity ($L_\mathrm{acc}$) for the adiabatic Schwarzschild (Sim. 1) and Kerr (Sim. 4) simulations shown in Figs.~\ref{fig:main-schwarzschild} and \ref{fig:main-kerr}, and described in Sections~\ref{sec:main-schwarzschild} and \ref{sec:main-kerr}. The spike at 11 hrs corresponds to the `pancake' shock generated from the stellar compression during the first pericentre passage. Accretion only begins at $\sim\,$$25$--$30$ hrs, when material reaches pericentre a second time. $L_\mathrm{acc}$ roughly traces $L_\mathrm{shock}$ once accretion begins. The bump in $L_\mathrm{shock}$ for the Schwarzschild case (left) corresponds to the first stream-stream collision. We show a $t^{-5/3}$ slope for comparison, although this fallback rate is only expected for parabolic orbits. Dotted line shows the Eddington limit.}
      \label{fig:main-lums}
   \end{center}
\end{figure*}

\section{Results} \label{sec:results}

\begin{figure*}
   \begin{center}
      \includegraphics[width=\textwidth]{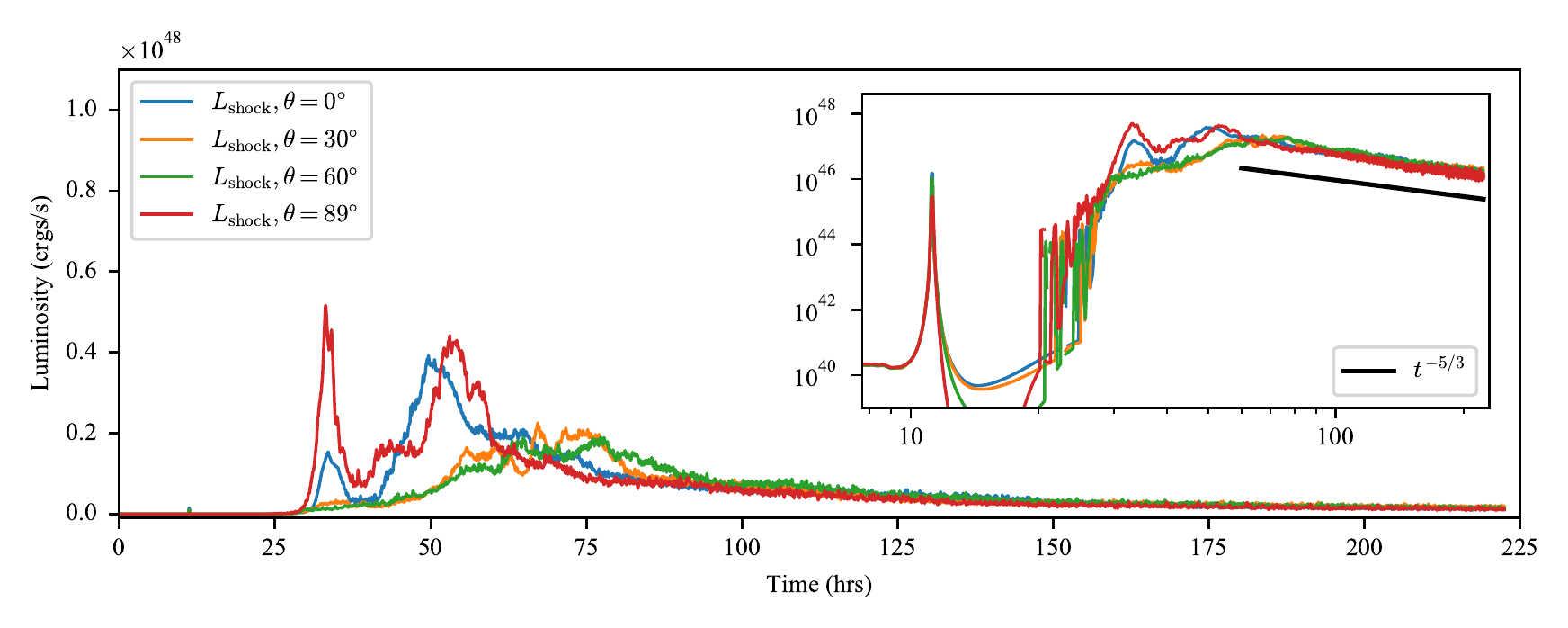}
      \includegraphics[width=\textwidth]{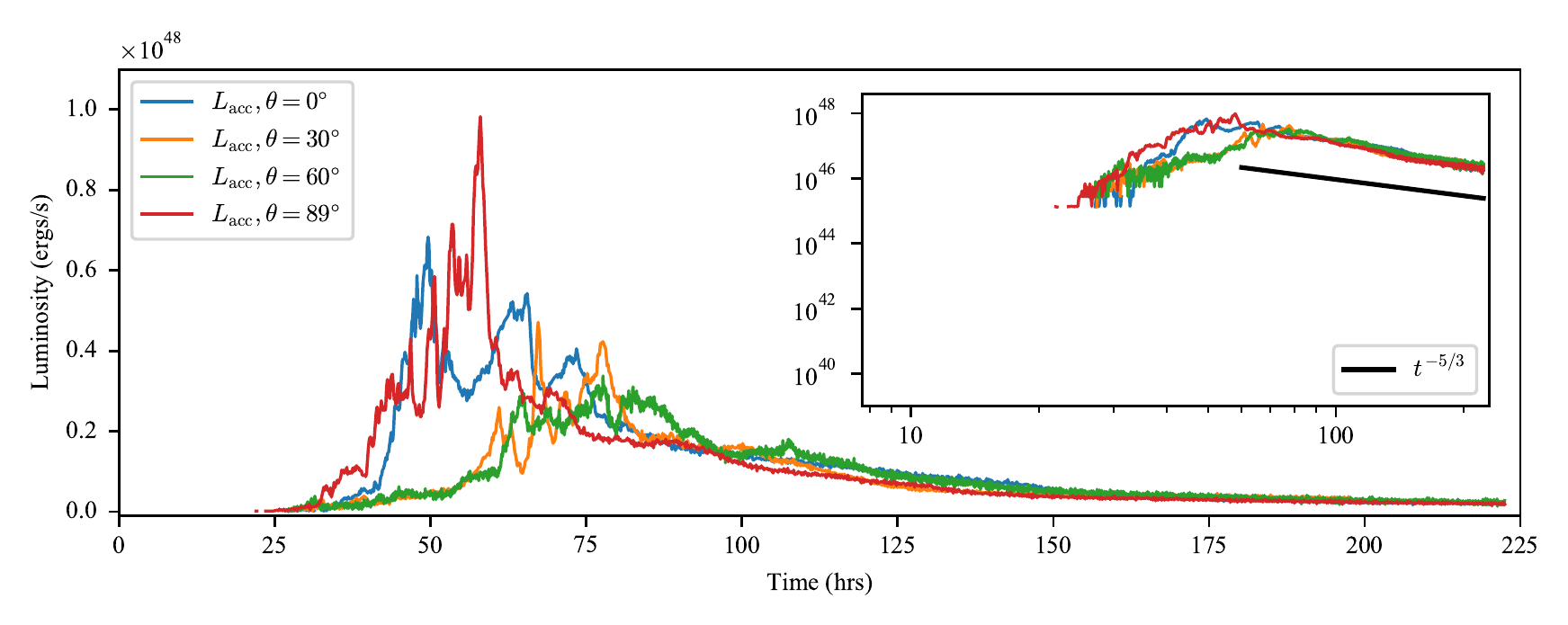}
      \caption{$L_\mathrm{acc}$ and $L_\mathrm{shock}$ for simulations 2, 3, 4 and 5 which differ only in their initial inclination angle. \textit{Top:} luminosity from viscous heating and shocks. \textit{Bottom:} luminosity from accretion. The same plots with log-log axes are shown as insets in each panel, along with a $t^{-5/3}$ slope for comparison.}
      \label{fig:inc-lums}
   \end{center}
\end{figure*}

\subsection{Disc formation around Schwarzschild black holes} \label{sec:main-schwarzschild}
Figure~\ref{fig:main-schwarzschild} shows snapshots of Simulation 1, in which the black hole is non-rotating and the initial stellar orbit lies in the $x$-$y$ plane with $e=0.95$ and $\beta=5$. This calculation is adiabatic.

As the star passes pericentre, relativistic apsidal precession shifts the orbit counter-clockwise by approximately $140\degree$. The orbital plane of the now disrupted star remains in the $x$-$y$ plane. On its return to the black hole, the stellar material is stretched into a long debris stream, with the head returning to pericentre at approximately 28.9hrs (first panel in Fig.~\ref{fig:main-schwarzschild}). Following this, the head of the stream intersects the tail due to relativistic apsidal precession (second panel). The intersection occurs close to pericentre since the degree of apsidal precession is large, which also means that the tangential velocity of the head is comparable to its component perpendicular to the tail. This produces a strong shock which is able to reduce a significant fraction of the stream's orbital energy, causing it to circularise and form a thick torus over the subsequent tens of hours. The torus extends to a radius of roughly $500R_g$ (approximately 5 au, or 1.4 times the initial apoapsis).

\subsection{Disc formation around Kerr black holes} \label{sec:main-kerr}
Figure~\ref{fig:main-kerr} shows snapshots of the disc formation process in Simulation 4. This calculation is the same as that shown in Figure~\ref{fig:main-schwarzschild} except the black hole is now rotating with spin $a=0.99$ and the initial stellar orbit has inclination $\theta=60\degree$.

The journey to the first pericentre passage is the same as in the equivalent Schwarzschild calculation. However after pericentre at 11.1hrs, nodal precession due to black hole spin moves the disrupted material into a different orbital plane, in addition to relativistic apsidal precession shifting the apocentre position in the prograde direction. The stellar debris after disruption is again stretched into a long stream, with the head returning to pericentre at approximately 28.9 hrs. This time nodal precession causes the stream to miss itself, shown at 33.4 hrs (second panel in Fig.~\ref{fig:main-kerr}).
However the head and tail of the stream still have a grazing encounter, since the stream has thickened with each pericentre compression, causing some material to begin circularising. During the third pericentre passage, the stream has thickened substantially and a low density cloud of material is already built up around the black hole. This causes all returning material to mildly shock and begin circularising. By 111 hrs most of the material has formed into a thick torus similar to the one formed around the Schwarzschild black hole, but inclined by approximately $60\degree$ to the $x$-$y$ plane. Between 100 and 220 hrs, the torus precesses about the black hole spin axis by roughly $4\degree$, corresponding to a precession period of 450 days.

\begin{figure*}
   \begin{center}
      \includegraphics[width=\textwidth]{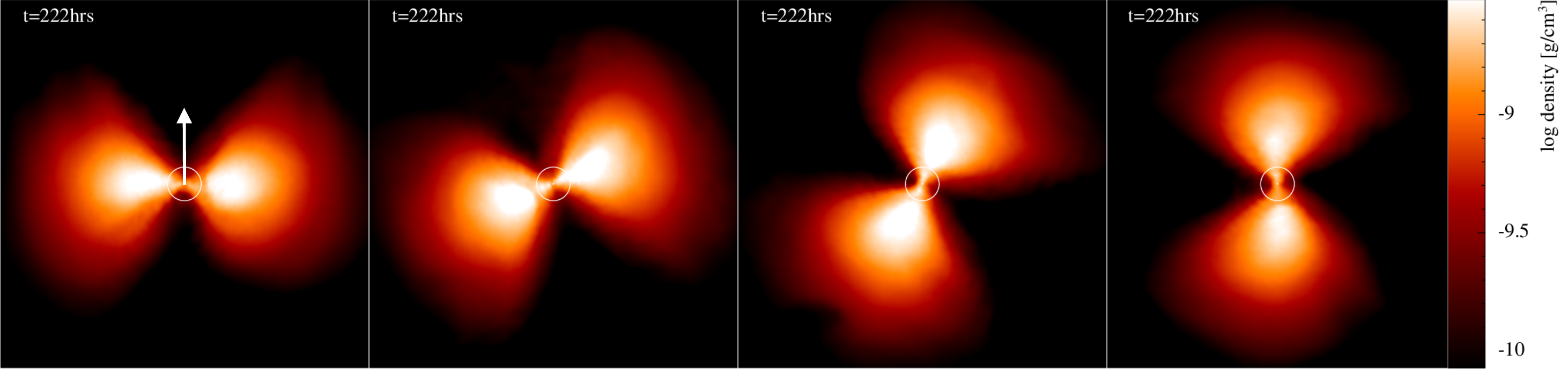}
      \caption{Cross sections of the final tori formed in simulations 2, 3, 4 and 5, taken through the planes defined by their total angular momentum vectors and the $z$-axis. Panels are 10 au $\times$ 10 au. The torus formed in each case has approximately the same inclination as the corresponding initial stellar orbit, i.e. $0\degree$, $30\degree$, $60\degree$ and $89\degree$ from left to right. We note that the angular momentum vector for the inclined calculations is `twisted' about the $z$-axis due to Lense-Thirring precession by approximately $37\degree$, $42\degree$ and $43\degree$ for inclinations $30\degree$, $60\degree$ and $89\degree$ respectively. The black hole spin vector is shown by the arrow, and white circles indicates the tidal radius $R_\mathrm{t}$.}
      \label{fig:inc-xsec}
   \end{center}
\end{figure*}

\subsection{Energy injection rate}
Figure~\ref{fig:main-lums} compares $L_\mathrm{acc}$ and $L_\mathrm{shock}$ in Simulations 1 and 4. The sharp spike of $\sim\,$$10^{46}$ erg s$^{-1}$ at approximately 11 hours corresponds to the heating generated by the pancake shock as the star passes through an effective nozzle at pericentre, which compresses the star in the direction perpendicular to its motion.
The non-rotating case shows a steeper rise than the rotating case, with a distinct peak at 33 hours in the shock luminosity, corresponding to the stream-self intersection. This peak is absent in the rotating case because the stream does not self-intersect at that time.
After 50 hours the energy injection rate reaches a maximum in the non-rotating case, then steadily declines as a power law. The maximum heating in the rotating simulation is reached slightly later at approximately 80 hours, before steadily declining. This turnover, in both cases, corresponds roughly to the disc/torus formation phase. We show the canonical $t^{-5/3}$ slope for comparison during the decay, although this trend is expected only for stars on initially parabolic orbits.
Once accretion begins, $L_\mathrm{acc}$ roughly traces $L_\mathrm{shock}$ throughout the rest of the simulation. This is not altogether surprising since $L_\mathrm{shock}$ includes both viscous heating and shock heating due to stream-stream collisions, which are the two processes that can drive accretion. Furthermore, the contribution to $L_\mathrm{shock}$ is dominated by particles at the accretion radius.
We discuss this further in Section~\ref{sec:discussion}.

\begin{figure}
   \begin{center}
      \includegraphics[width=\columnwidth]{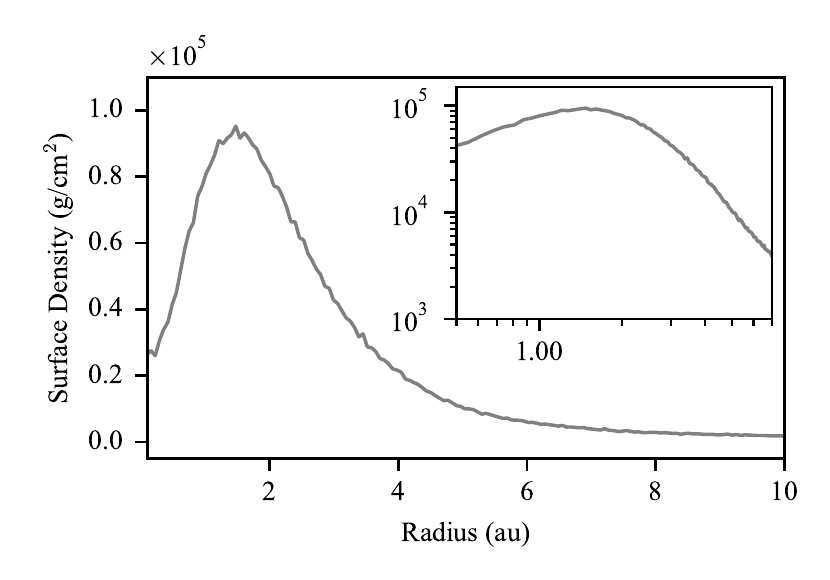}
      \caption{Surface density as a function of radius for Simulation 2. Inset shows the same plot with log-log axes. We measure the surface density profile to be roughly $\propto r^{-2.9}$ in the outer part of the disc.}
      \label{fig:sigma}
   \end{center}
\end{figure}

\subsection{Misalignment angle}
Figure~\ref{fig:inc-lums} compares $L_\mathrm{acc}$ and $L_\mathrm{shock}$ between simulations 2, 3, 4 and 5, which vary only in their inclinations, $0\degree$, $30\degree$, $60\degree$ and $89\degree$, respectively, around a black hole with spin $a=0.99$.
For moderate inclinations ($30\degree$ and $60\degree$) the tidal stream avoids the first self-intersection episode that occurs at $\sim\,$$33$ hrs in the other two simulations, shown by the spikes in $L_\mathrm{shock}$ at that time. However at the largest inclination of $89\degree$, we find that nodal precession does not prevent this first stream-stream collision.

We also find that the $0\degree$ and $89\degree$ simulations are brighter by approximately an order of magnitude during the early stages of the calculation, and form discs at $\sim\,$75 hrs, roughly 1 orbital period earlier than the $30\degree$ and $60\degree$ calculations.

For increasing inclinations the stellar orbits penetrate fractionally deeper leading to a larger degree of both apsidal and nodal precession. The deeper encounters also produce stronger shocks during the pericentre compression, which output more heat and thus cause thicker debris streams. For the $89\degree$ inclined orbit, the combination of these effects leads to a stream self-intersection episode that is brighter than even the $0\degree$ case, which suggest that stream-stream collision behaviour is not a simple function of inclination.

\begin{figure*}
   \begin{center}
      \includegraphics[width=\textwidth]{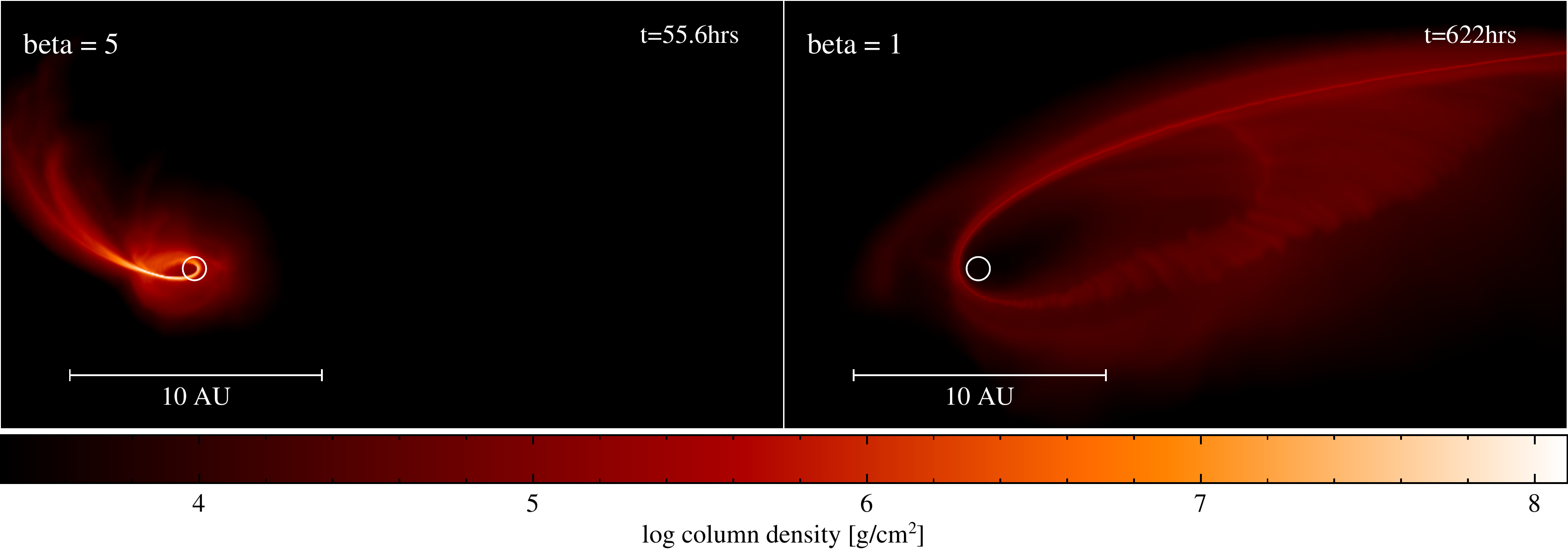}
      \caption{Snapshots of Simulations 4 and 6 after 2.5 orbital periods each, with penetration factors $\beta=5$ and $\beta=1$ (\textit{left} and \textit{right} panels respectively), projected in the initial orbital plane. The stream in the $\beta=5$ simulation intersects close to the black hole, while for $\beta=1$ it intersects at apoapsis where the stream has a low density and low velocity. White circle indicates the tidal radius $R_\mathrm{t}$.}
      \label{fig:beta-render}
   \end{center}
\end{figure*}

\subsection{Remnant disc properties}
Figure~\ref{fig:inc-xsec} shows cross sections of the discs/tori formed at the end of calculations 2, 3, 4 and 5. We find that each disc has approximately the same inclination as its corresponding initial stellar orbit, and that more inclined orbits have their angular momentum vectors `twisted' about the $z$-axis by a larger degree due to Lense-Thirring precession.

The torus has a similar structure in each case. Since the gas cannot cool in these simulations the tori remain extended beyond the tidal radius, extending to a radius of roughly $500R_g$, or approximately 5 au, with an aspect ratio $H/R \sim 1$.
Figure~\ref{fig:sigma} shows the surface density profile of Sim. 2 at $t=222$ hrs, computed in the manner of \citet{lodatoprice10}. That is, we bin particles by cylindrical radius into annuli, take the mass of each annulus and then divide by their cross-sectional area. The surface density has a peak of $\approx9\times10^4$ g cm$^{-2}$ at $r\approx1.5$ au, and decays as a power law roughly $\propto r^{-2.9}$ in the outer part of the disc.

\begin{figure}
   \begin{center}
      \includegraphics[width=\columnwidth]{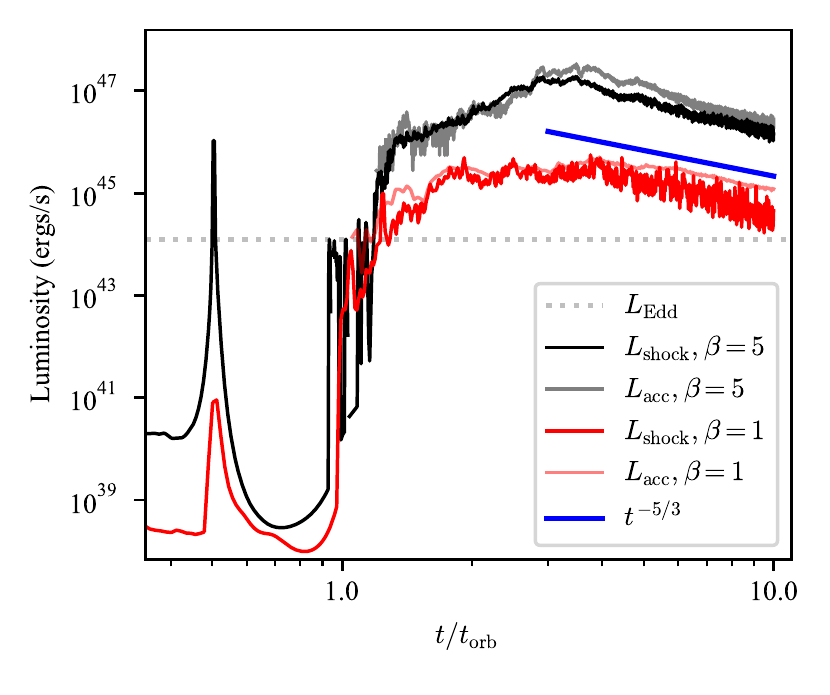}
      \caption{$L_\mathrm{acc}$ and $L_\mathrm{shock}$ for simulations 4 and 6 which differ only in their penetration factor $\beta$. Although the orbital period $t_\mathrm{orb}$ for $\beta=1$ is approximately 11 times that of $\beta=5$, their behaviour is similar as a function of the number of orbits. The energy injection rate is approximately 2 orders of magnitude greater for $\beta=5$ throughout the simulation. Dotted line shows the Eddington limit.}
      \label{fig:beta-lums}
   \end{center}
\end{figure}

\begin{figure*}
   \begin{center}
      \includegraphics[width=\textwidth]{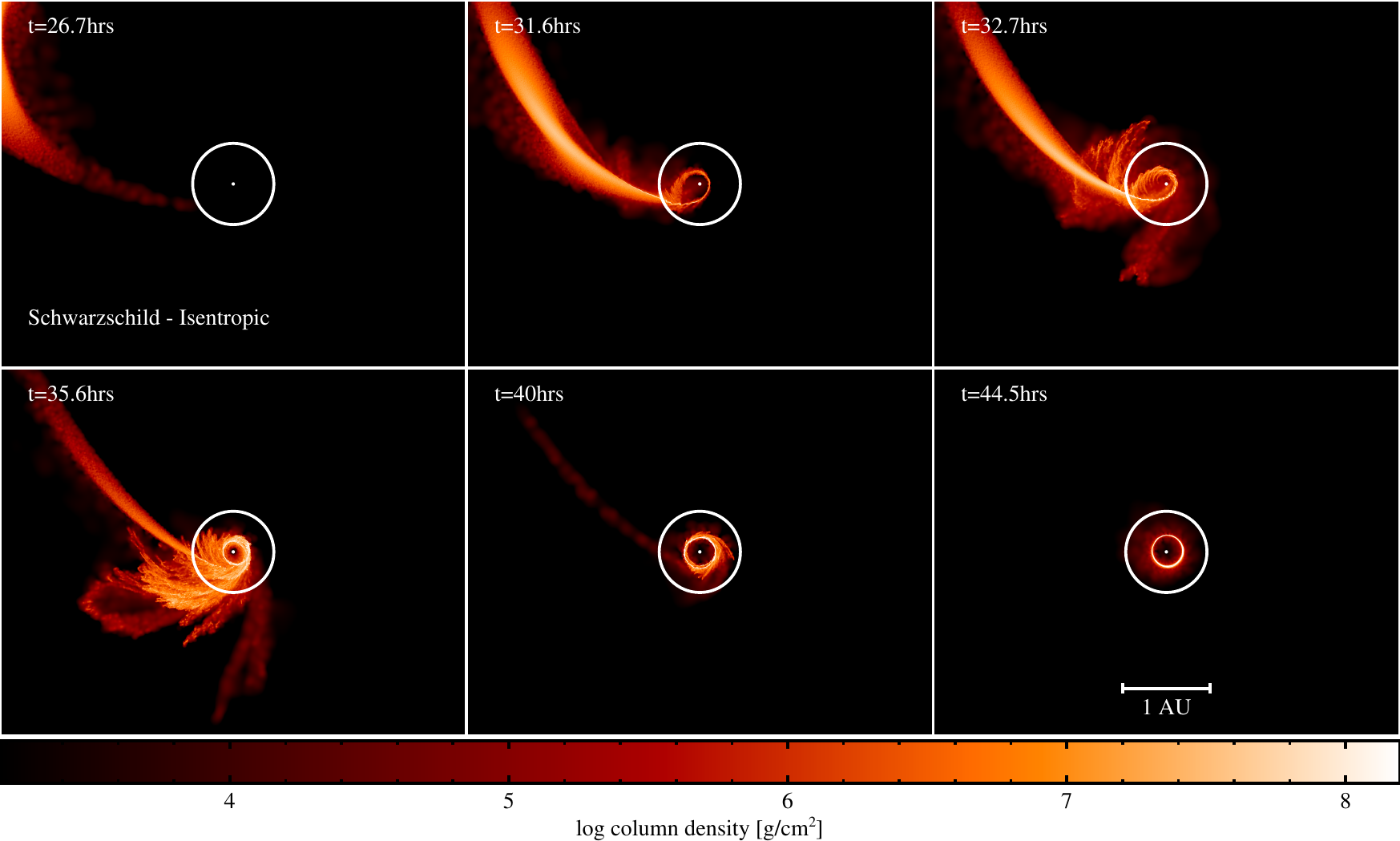}
      \caption{Snapshots of the debris fallback projected in the $x$-$y$ plane for the $e=0.95$, $\beta=5$, $\theta=0\degree$, $a=0$ isentropic simulation (Sim. 7). The debris stream remains much thinner than in the corresponding adiabatic calculation (Fig.~\ref{fig:main-schwarzschild}) since energy from viscous and shock heating is discarded. A very narrow ring is formed quickly after self-intersection, at twice the periapsis radius. White circle indicates the tidal radius $R_\mathrm{t}$.}
      \label{fig:isentropic-schwarzschild}
   \end{center}
\end{figure*}

\begin{figure*}
   \begin{center}
      \includegraphics[width=\textwidth]{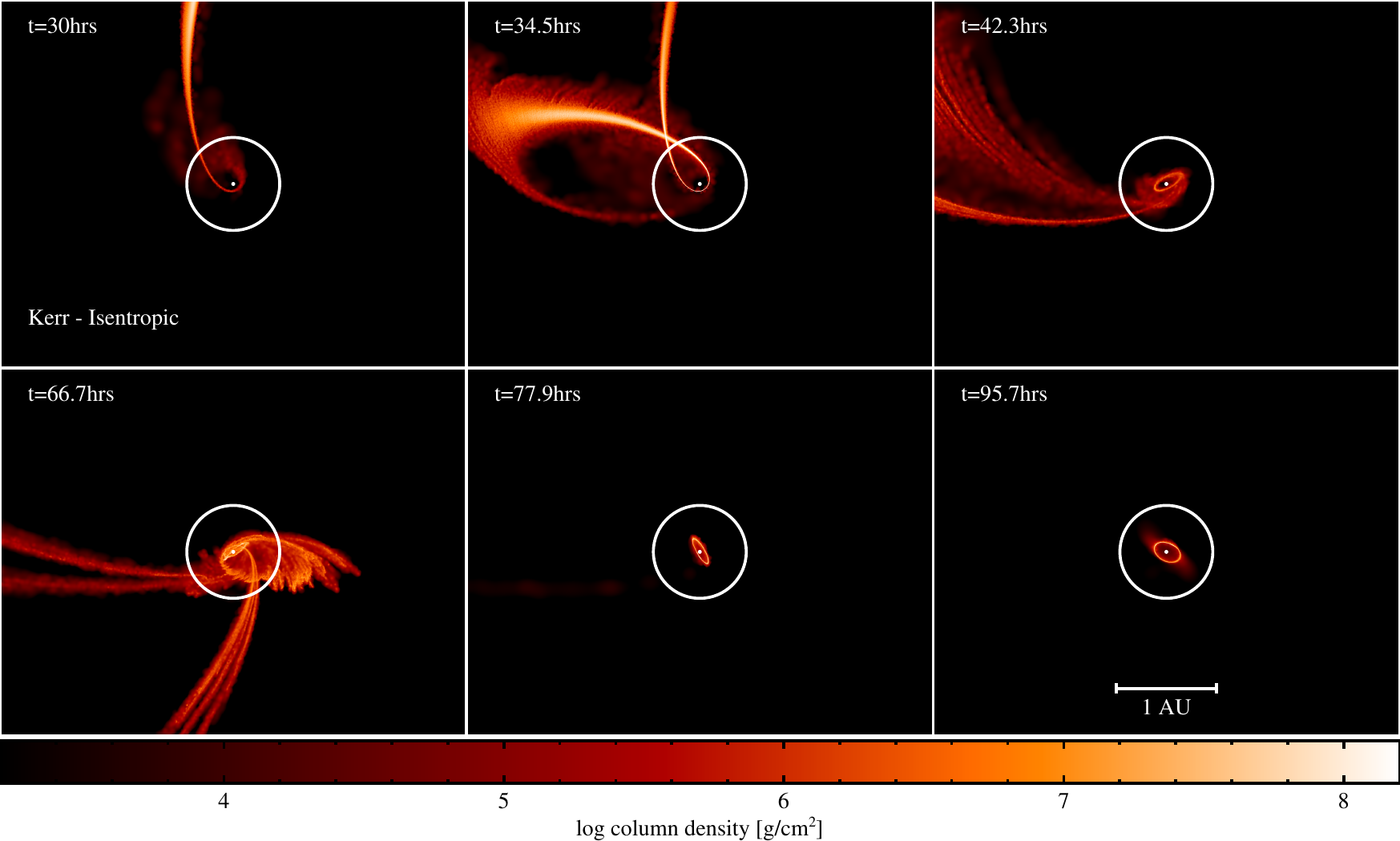}
      \caption{Snapshots of the debris fallback projected for the $e=0.95$, $\beta=5$, $\theta=60\degree$, $a=0.99$ isentropic simulation (Sim. 8), projected in the initial orbital plane. The stream avoids self-intersection after one orbit, but over subsequent orbits the stream circularises into a narrow ring at $\sim\,$$1.6R_\mathrm{p}$ inclined by $\sim\,$$50\degree$ to the black hole spin plane. The ring precesses about the black hole spin axis, completing $\sim\,$4 revolutions between approximately 70 -- 110 hrs, with a period of $\sim\,$10 hours. White circle indicates the tidal radius $R_\mathrm{t}$.}
      \label{fig:isentropic-kerr}
   \end{center}
\end{figure*}

\subsection{Penetration factor}
Figure~\ref{fig:beta-render} compares snapshots of simulations 4 and 6 with $e=0.95$, $a=0.99$ and $\theta=60\degree$, after 2.5 orbits each, where only the penetration factor $\beta$ is varied between the two calculations.
Figure~\ref{fig:beta-lums} shows $L_\mathrm{acc}$ and $L_\mathrm{shock}$ as a function of the number of orbital periods.
The orbital period for $\beta=1$ is $5^{3/2}\approx11$ times that of $\beta=5$, thus disc formation happens much more rapidly for higher $\beta$. However, they both take approximately the same number of orbits ($\sim\,$$4$ -- 5) to circularise.

In the deep encounter, the tidal stream undergoes a higher degree of both apsidal and nodal precession. This causes the stream to have an eventual self-intersection at a few periapsis radii from the black hole, and to form a torus in a plane that is different to the initial orbital plane (but with the same inclination). The stellar debris for the $\beta=1$ case however remains largely in the same orbital plane in which it started. Furthermore, it self-intersects almost at apoapsis, where the relative velocities between the head and tail are small and the debris is tenuous. As a consequence, the heating rate is 2 orders of magnitude fainter over the course of the simulations, and the final torus is $\sim\,$$5$ times greater in size than the $\beta=5$ case, extending to a radius of roughly $2500R_g$ (approximately 25 au, or 1.4 times the initial apoapsis).

The radius of self-intersection in both cases is roughly consistent with the analytical prediction of \citet{bonnerotrossilodato17}, but is difficult to measure accurately since the stream spreads after its periapsis passage.

\subsection{Radiative efficiency} \label{sec:isentropic}
To investigate the influence of cooling on the disc formation process we repeated the simulations from Section~\ref{sec:main-schwarzschild} and \ref{sec:main-kerr} but with an isentropic approximation as described in Section~\ref{sec:cooling}. This corresponds to efficient radiative cooling, while still allowing for (reversible) heating due to compressive $P\mathrm{d}V$ work.

\begin{figure*}
   \begin{center}
      \includegraphics[width=\columnwidth]{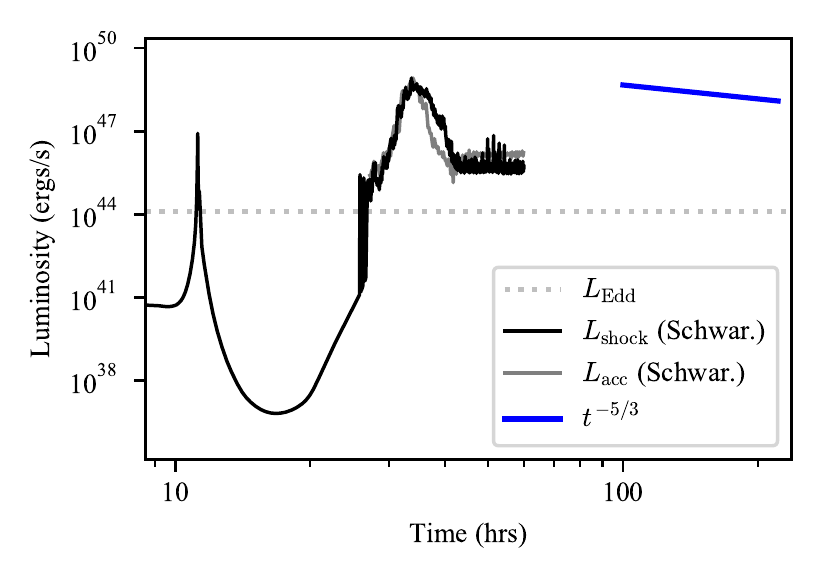}
      \includegraphics[width=\columnwidth]{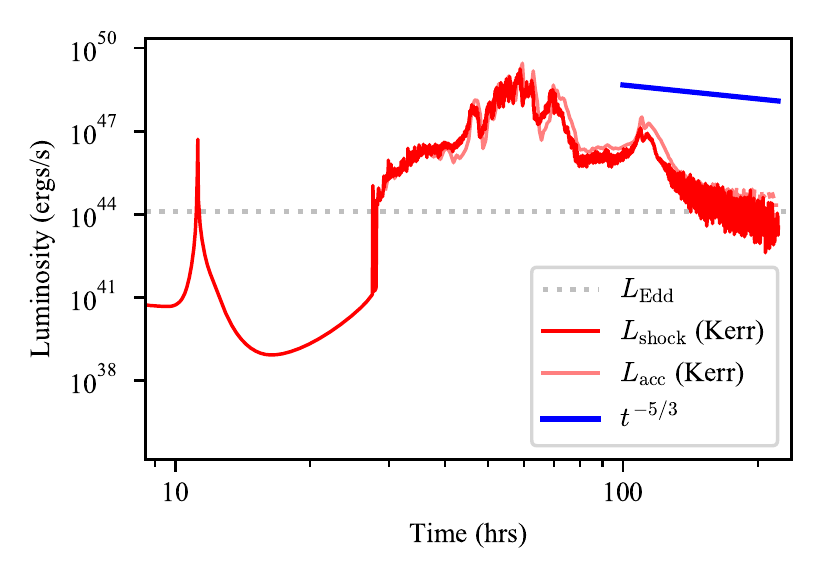}
      \caption{$L_\mathrm{acc}$ and $L_\mathrm{shock}$ for the isentropic Schwarzschild (Sim. 7) and Kerr (Sim. 8) simulations shown in Figs.~\ref{fig:isentropic-schwarzschild} and \ref{fig:isentropic-kerr}, and described in Section~\ref{sec:isentropic}. The peak luminosities are roughly 2 orders of magnitude greater than the corresponding adiabatic counterparts shown in Fig.~\ref{fig:main-lums}. Dotted line shows the Eddington limit.}
      \label{fig:isentropic-lums}
   \end{center}
\end{figure*}

Figure~\ref{fig:isentropic-schwarzschild} shows the evolution of Simulation 7 in which the black hole is non-rotating. Figure~\ref{fig:isentropic-kerr} shows the equivalent calculation around a Kerr black hole with spin $a=0.99$ and initial orbital inclination $\theta=60\degree$ (Simulation 8). Since the heat generated from viscosity and shocks is no longer contained within the gas, the tidal streams in the isentropic calculations remain thinner than in the adiabatic ones. They also form a small, thin ring at the end of the calculation close to the periapsis radius, in contrast to the large, thick tori produced by the adiabatic calculations. The peak energy injection and accretion luminosities produced by the isentropic calculations are also roughly 2 orders of magnitude brighter than their adiabatic counterparts, as shown in Figure~\ref{fig:isentropic-lums}.

In the Schwarzschild case there is one self-intersection, which very quickly circularises nearly all of the material into a thin ring in the $x$-$y$ plane by $\sim\,$$43$ hrs at a radius  approximately twice the pericentre distance. This is consistent with the circularisation radius $r_\mathrm{c}=(1+e)r_\mathrm{p}$ predicted from the conservation of angular momentum for Keplerian orbits \citep{bonnerotrossilodato16}. At this point the black hole enters a state of roughly constant accretion, shown by the plateau in the left panel of Figure~\ref{fig:isentropic-lums}.

In the Kerr simulation a direct self-intersection is avoided but a grazing encounter between the head and tail still causes some material to lose orbital energy and build up close to the black hole, similar to the adiabatic calculations. As the rest of the stream returns to pericentre, transient rings of material are created and destroyed between $\sim\,$$40$--$70$ hrs, approximately 8 times. This is reflected in $L_\mathrm{acc}$ and $L_\mathrm{shock}$ (Fig.~\ref{fig:isentropic-lums} \textit{right}) which take longer to reach a maximum than the Schwarzschild simulation and show multiple peaks between $40$--$70$ hrs. Following this, most of the remaining material collects into a narrow ring inclined by $\sim\,$$50\degree$ to the black hole spin plane with a radius equal to roughly $\sim\,$1.6 times the pericentre distance. Over the course of the next 40 hours the ring precesses with a period of approximately 10 hours, completing 4 revolutions before being completely accreted by the black hole, shown by the final spike in $L_\mathrm{acc}$ and $L_\mathrm{shock}$ at $\sim\,$$110$ hrs.

\section{Discussion} \label{sec:discussion}
We presented eight numerical GRSPH simulations testing the disc formation process from the tidal disruption of stars by rotating supermassive black holes. \citet{hayasakistoneloeb13} were the first to demonstrate circularisation of the debris streams and formation of an accretion disc in numerical simulations. They showed that the main cause of circularisation is relativistic apsidal precession leading to self-intersection of the debris stream. \citet{bonnerotrossilodato16} confirmed this using an improved pseudo-Newtonian potential, and further showed that the structure of the resultant disc depends on the cooling efficiency. Both studies performed SPH calculations in pseudo-Newtonian potentials and were not able to simulate spinning black holes. In our Schwarzschild metric calculations, using General Relativistic SPH without pseudo-Newtonian approximations, we confirm this basic picture --- namely that relativistic apsidal precession leads to rapid debris circularisation.

\citet{guillochonramirez-ruiz15} suggested that black hole spin would ruin this picture, since Lense-Thirring precession causes the debris stream to change plane and thereby avoid self-intersection (see also \citealt{hayasakistoneloeb16}). Figure~\ref{fig:main-kerr} shows that the stream indeed misses itself at the second pericentre passage (at $t=33$ hrs), however, this is not fatal to disc formation because the stream does not miss by much. This is because nodal (spin-induced) precession is weak compared to apsidal (Schwarzschild) precession, meaning that the stream crossing occurs before nodal precession has time to move it significantly out of the plane. Hence the stream misses by only a few degrees. By the next orbit the stream is already wide enough to successfully self-intersect and circularise, in contrast to the many dark windings of the stream implied by \citet{guillochonramirez-ruiz15}. The `dark year' proposed by these authors is therefore bright. The caveat here is that we considered only orbits that were prograde with respect to the black hole spin, whereas \citet{hayasakistoneloeb16} considered retrograde orbits and found that circularisation could be more substantially delayed.

We found that disc formation --- at least for bound orbits --- appears to be robust with regards to black hole spin and inclination. Although nodal precession of the debris stream's orbital plane can prevent a direct self-intersection, the head and tail typically do not miss by much and still have a grazing encounter owing to their non-negligible cross-sections. As a consequence, the gas is shocked and rapidly circularises to form a thick torus as it does when the black hole is not rotating. However, the peak energy injection rate tends to be slower, and the rise time longer, than if the stream has a direct self-intersection.

As in \citet{bonnerotrossilodato16}, we found that the structure of the remnant disc depends strongly on the cooling efficiency. With no cooling (i.e. the adiabatic calculations), the debris settles into a thick and extended torus, while for radiatively efficient cooling (i.e. the isentropic calculations) a narrow ring is formed at a radius slightly greater than the pericentre distance. In both cases the remnant disc precesses about the black hole spin axis if it is misaligned.
We estimated a precession period of approximately 10 hours for the ring and approximately 450 days for the torus, in the cases where $\theta=60\degree$, $a=0.99$, $\beta=5$ and $e=0.95$.
The global precession period predicted by \citet{franchinilodatofacchini16} for a rigidly precessing torus is between approximately 2 and 4 days around a $10^6M_\odot$ black hole with spin $a=0.99$. However the disc properties considered by these authors differ from those of our remnant discs. Our discs extend to $\sim\,5$ au when $\beta=5$, compared to their assumption of $R_\mathrm{out}=2R_\mathrm{t}$ corresponding to $\approx1$ au. Since Lense-Thirring precession frequency scales as $r^{-3}$, this corresponds to a factor of $\sim\,$125 difference in precession timescale, which explains the discrepancy. The precession rate of the ring formed in our isentropic calculation is consistent with the Lense-Thirring rate at its radius ($\approx15 R_g$).

The peak luminosity and rise time of $L_\mathrm{acc}$ and $L_\mathrm{shock}$ depend strongly on the penetration factor $\beta$. This is because the orbital period is much shorter for orbits with high $\beta$, and the apsidal precession frequency is higher. As a consequence the stream self-intersects at a radius close to pericentre, leading to stronger shocks and faster circularisation. For deep $\beta=5$ encounters, the typical peak luminosities are on the order of $\sim\,$$10^{47}$ erg s$^{-1}$, taking several tens of hours to reach this peak after the first pericentre passage. This is roughly 3 orders of magnitude greater than the Eddington luminosity.
In contrast, $L_\mathrm{acc}$ and $L_\mathrm{shock}$ in the $\beta=1$ calculation are 2 orders of magnitude lower than with $\beta=5$, and take $\sim\,$11 times longer to reach their peak. Thus the relative penetration depth may be inferrable from TDE observations.

Our remnant discs are larger than those shown in \citet{bonnerotrossilodato16} by a factor of $\sim$2--3 ($\approx5R_\mathrm{t}$ compared to their $\approx2R_\mathrm{t}$, see their Fig. 8). This occurs because we consider a larger eccentricity of $e=0.95$ compared to their calculation which used $e=0.8$, implying greater orbital energy dissipation from the circularisation process. We confirmed that our $e=0.8$ calculations produced similar disc sizes as those shown in \citet{bonnerotrossilodato16}. This suggests that even larger and hotter remnants would be produced in the parabolic case.

The initial pericentre passage is accompanied by a sharp spike in the energy injection rate ($L_\mathrm{shock}$), corresponding to the shocks produced by the `pancaking' of the star during its periapsis compression. This has been predicted to be observable as a brief X-ray shock breakout with a peak luminosity between $\sim\,$$10^{40}$--$10^{43}$ erg s$^{-1}$ \citep{kobayashilagunaphinney04,guillochonramirez-ruizrosswog09,yalinewichguillochonsari19}, depending on the amount of energy lost as the shock moves out through the star. For $\beta=5$, our calculations produce a spike of $\sim\,$$10^{46}$ erg s$^{-1}$, $3$--$6$ orders of magnitude greater than the predicted X-ray luminosity, although this is only the energy dissipation rate and not a true luminosity.

Most previous studies estimated the luminosity evolution simply from the predicted mass return rate (see studies that do not follow fallback in Section~\ref{sec:intro}) or based on other assumptions about the dynamics of the stream's evolution \citep[e.g.][]{bonnerotrossilodato17}. We computed the energy injection rate from first principles, based on either the irreversible energy dissipation rate or the accretion rate at $5R_g$ in our simulations. This approach is similar to \citet{shiokawakrolikcheng15} who estimated the heating rate by integrating the divergence of the heat flux in their calculations, however they were unable to distinguish between heating from reversible adiabatic compression and true entropy generation.

The evolution of $L_\mathrm{shock}$ shows features that may be identified with the energy release caused by the `pancake shock', stream-stream collisions and viscous accretion (see Figs.~\ref{fig:main-lums}, \ref{fig:inc-lums}, \ref{fig:beta-lums} and \ref{fig:isentropic-lums}). We found that the luminosity from shocks and viscous heating $L_\mathrm{shock}$ roughly traces the accretion luminosity $L_\mathrm{acc}$ in all our calculations. This is expected since $L_\mathrm{shock}$ does not distinguish between viscous dissipation and shock heating, which are the only two mechanisms available to reduce orbital energy and drive accretion. One might expect this to be a sign that material is being accreted ballistically \citep[e.g.][]{svirskipirankrolik17,bonnerotrossilodato17}, however this is only true if the energy release is dominated by shocks at the location of stream self-intersection. In our calculations the heating rate is dominated by regions close to the accretion radius, suggesting that material is being accreted viscously.

A major caveat to our approach is that our simulations do not include radiation transport. We instead perform calculations in which the gas is either completely adiabatic or isentropic. Although these are two extreme regimes --- since adiabatic calculations trap all heat while the isentropic calculations instantly remove heat produced by shocks and viscous heating --- they provide upper and lower bounds to the true behaviour.
Since the energy generation rate is super-Eddington, most of the energy cannot be transported as radiation and must instead be transported via mechanical outflows. This is precisely what we find in the adiabatic calculations. Our remnants in this case are likely closer to the truth, in the sense that the bulk of the energy goes into gas heating and kinematics, and the gas is not able to radiatively cool.
Furthermore, any photons radiated in shocks would take a long time to diffuse out through optically thick material \citep[e.g.][]{loebulmer97,rothkasenguillochon16}, and be reprocessed into different wavelengths, possibly carrying information about the outflow geometry \citep[e.g.][]{nichollblanchardberger19,leloudasdaiarcavi19}.

The energy injection and accretion rates we calculate are only approximations to the true light curve, particularly in the adiabatic case where $L_\mathrm{shock}$ represents the energy generation rate, and not the radiated luminosity. For the isentropic calculations however, discarded shock heating does genuinely reflect energy emitted from the gas. 
Our $L_\mathrm{acc}$ simply tracks the mass flux at $r=5R_g$, which may either produce radiation or simply be advected into the black hole. Computing the true luminosity evolution is beyond the scope of this paper, since it would require the development of a General Relativistic radiation hydrodynamics scheme \citep[e.g.][]{daimckinneyroth18}.

We also do not include magnetic fields in our calculations, which are thought to be important in the dynamics of TDEs \citep{bonnerotrossilodato17,bonnerotpricelodato17,svirskipirankrolik17}. In particular, the magneto-rotational instability (MRI) is thought to be the mechanism driving disc accretion. In our calculations disc accretion is driven by numerical viscosity, which is resolution dependent --- in particular it is higher at lower resolutions. As a consequence, the black hole accretion rate may be overestimated and thus the lifetime of the disc underestimated (see Appendix~\ref{sec:appendix}).

Since the parameter space is already large for this problem, we focused on only prograde orbits. We also did not consider stellar spin \citep{golightlycoughlinnixon19,sacchilodato19}, or stellar structure \citep{lodatokingpringle09}, both of which have been shown to affect the fallback rate. The largest caveat of all is that we restricted our simulations to bound orbits.

\section{Conclusion} \label{sec:conclusion}
We presented general relativistic numerical simulations of the tidal disruption of stars on bound orbits by SMBHs. Our calculations self-consistently employed relativistic hydrodynamics in the Kerr metric along with energy injection rates calculated from the irreversible dissipation in the gas and the mass accretion rate onto the black hole.
Our key findings are as follows:
\begin{enumerate}
   \item Nodal precession due to black hole spin does not prevent disc formation in TDEs, merely causing a short delay. 
   \item Radiatively efficient cooling produces a narrow ring of orbiting material, while no cooling produces an extended, thick torus. Lense-Thirring precession causes the remnant disc in both cases to precess about the black hole spin axis.
   \item Thick discs/tori can be formed robustly with the same inclination as the initial stellar orbit. They have an aspect ratio of $\sim\,$1, surface density profiles roughly $\propto r^{-3}$, and extend to of order the apoapsis radius.
   \item TDEs with high penetration factors are significantly more luminous and produce discs faster than those with a lower penetration factor. We find $\beta=5$ TDEs to be $\sim\,$2 orders of magnitude higher energy injection rates than those with $\beta=1$.
\end{enumerate}
Whether or not these results extend to stars disrupted on parabolic orbits remains to be investigated.

\section*{Acknowledgements}
We thank Kimitake Hayasaki, Paul Lasky and Morgan MacLeod for useful discussions and feedback. We acknowledge CPU time on OzSTAR funded by Swinburne University and the Australian Government. DL is funded through the Australian government RTP (Research Training Program) stipend. DJP is grateful for funding from the Australian Research Council via FT130100034. This project has received funding from the European Union's Horizon 2020 research and innovation programme under the Marie Sklodowska-Curie grant agreement No 823823 (Dustbusters RISE project).

\bibliographystyle{mnras}
\bibliography{dave}
\appendix
\section{Effect of resolution} \label{sec:appendix}
\begin{figure}
   \begin{center}
      \includegraphics[width=\columnwidth]{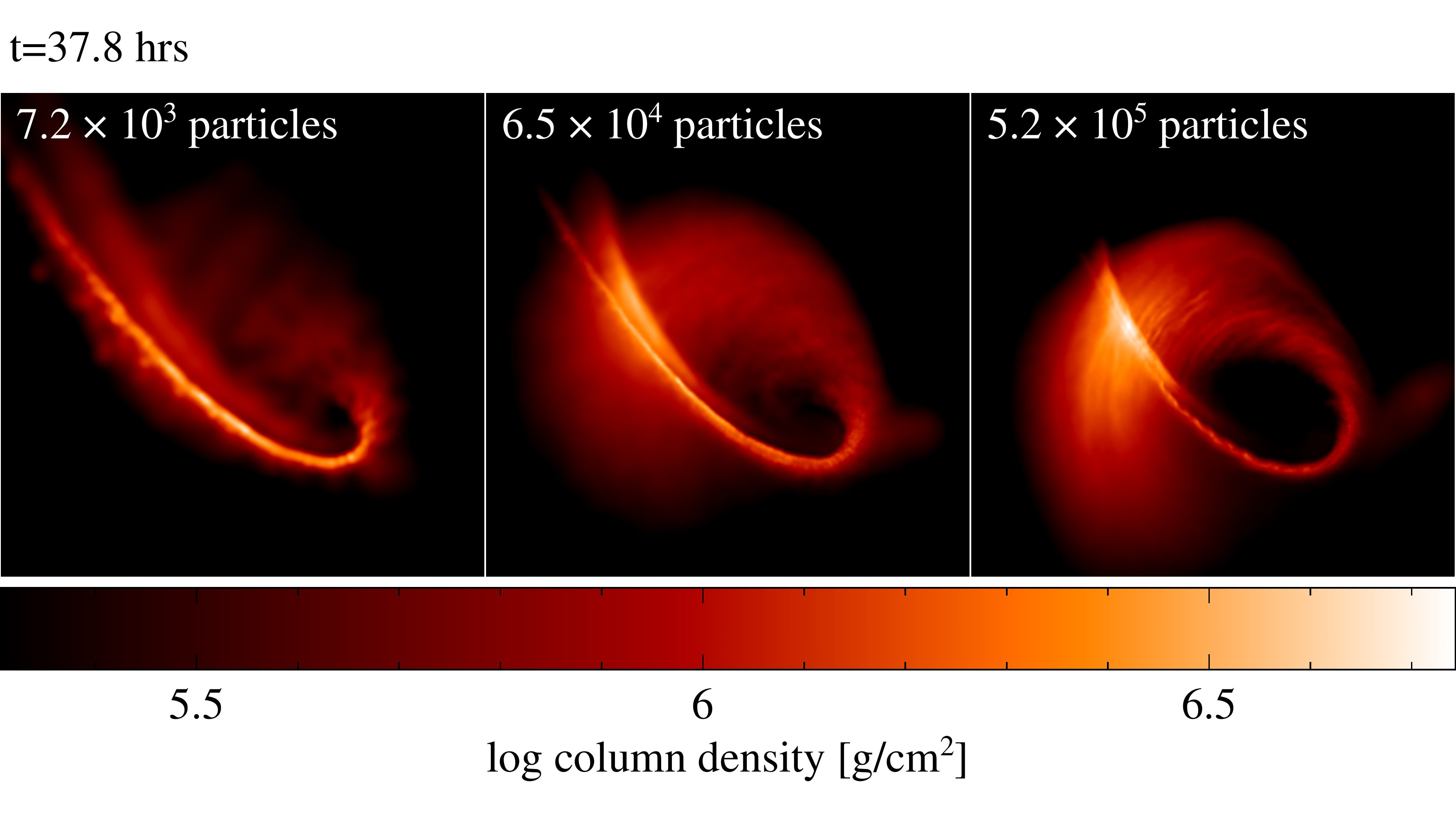}
      \includegraphics[width=\columnwidth]{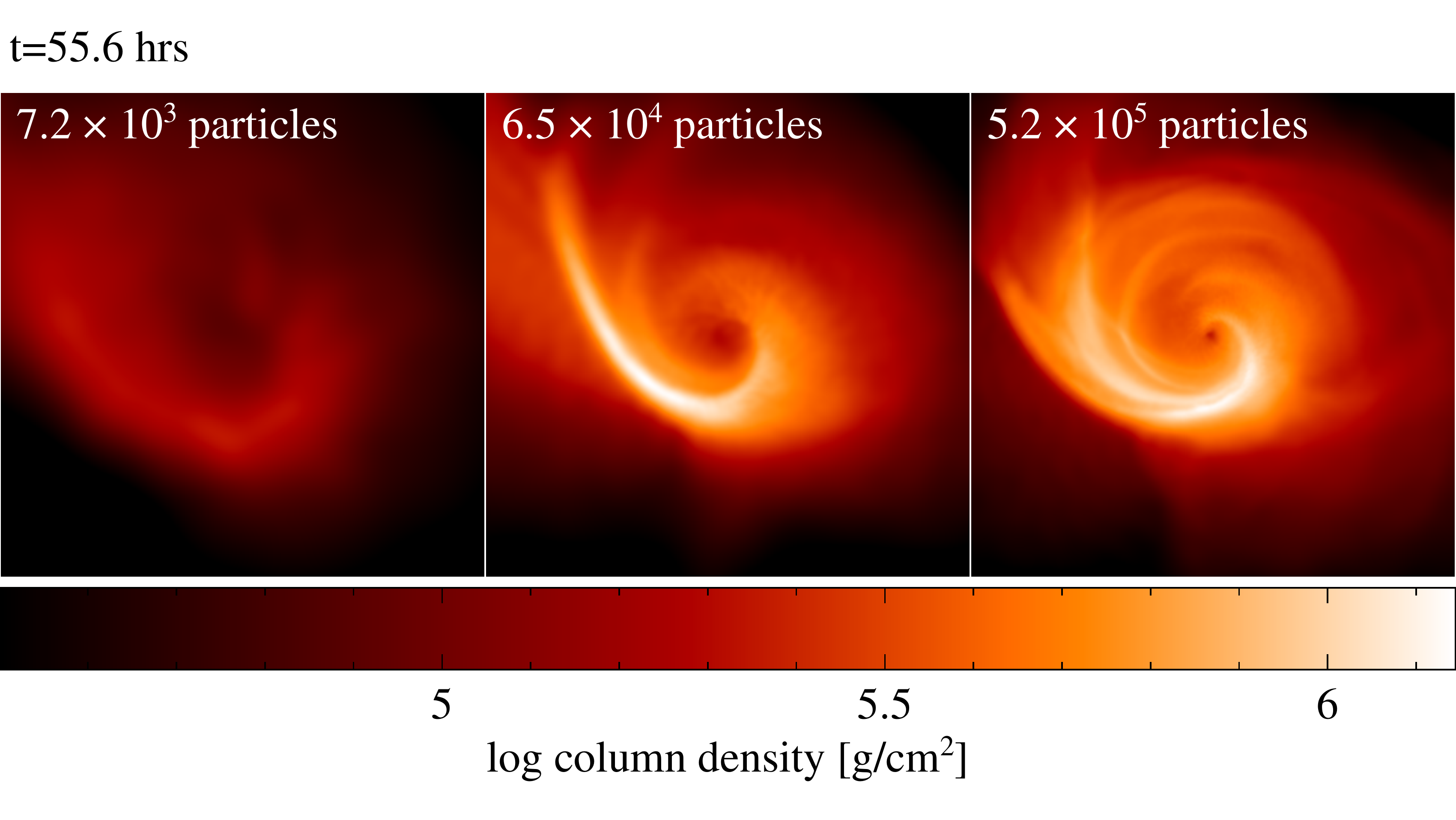}
      \includegraphics[width=\columnwidth]{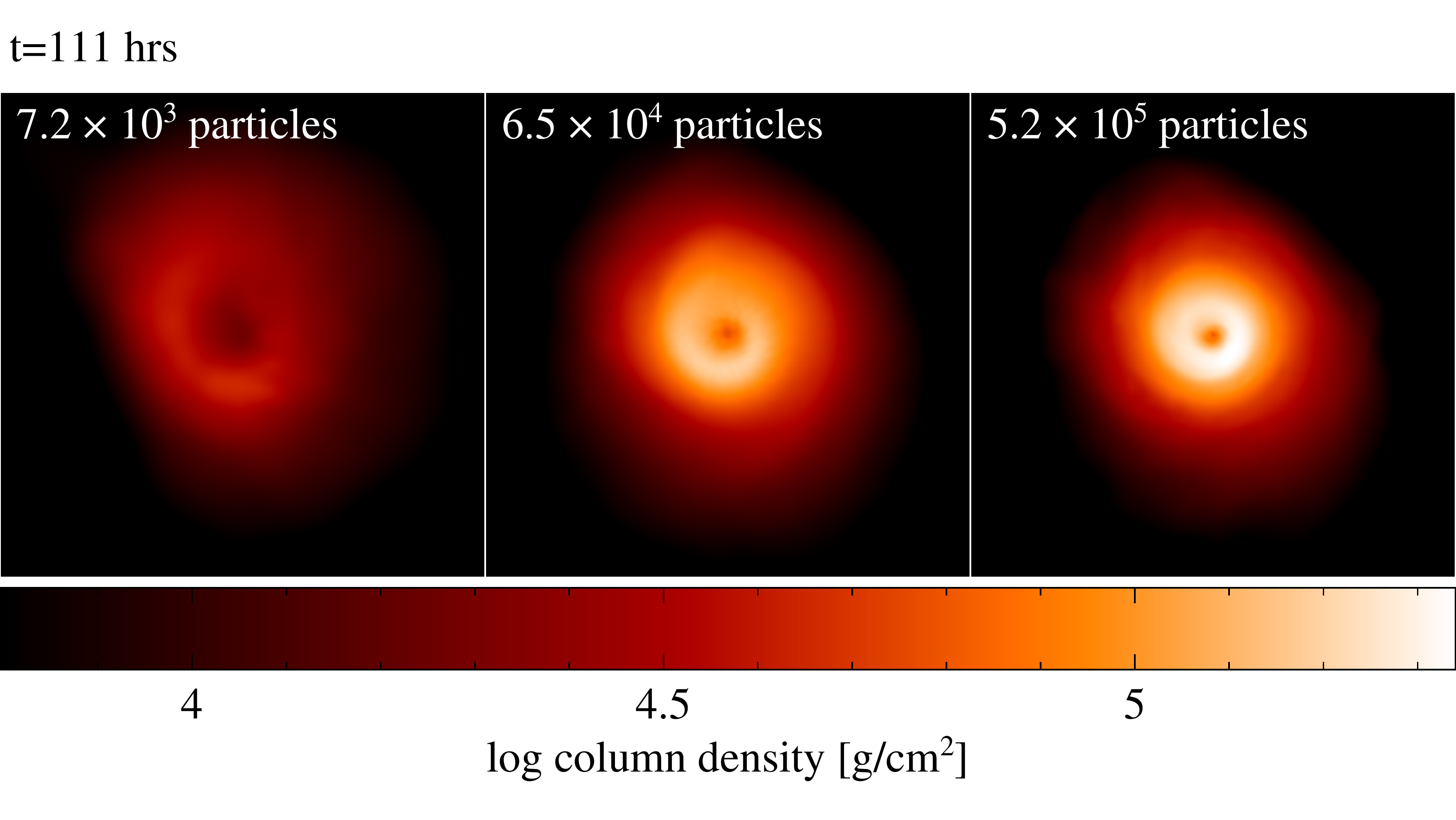}
      \caption{Effect of resolution in Simulation~1. \textit{Top row:} comparison of the self-intersection at $t=37.8$ hrs. \textit{Middle row:} comparison of the accretion dynamics at $55.6$ hrs. \textit{Bottom row:} comparison of the remnant torus at $t=111$ hrs. Total number of particles increases by approximately a factor of 8 between panels from left to right, with the right panel corresponding to the resolution shown in Fig.~\ref{fig:main-schwarzschild}.} 
      \label{fig:resolution}
   \end{center}
\end{figure}

Figure~\ref{fig:resolution} shows the effect of resolution in Simulation~1 at three different resolutions. The \textit{top} row compares the self-intersection at $t=37.8$ hrs, the \textit{middle} row shows the accretion dynamics at $t=55.6$ hrs, and the \textit{bottom} row compares the remnant torus formed at $t=111$ hrs. 
We increase the number of particles by approximately a factor of 8 between each panel from left to right, corresponding to a factor of 2 increase in spatial resolution.
The self-intersection is better resolved and the torus is more defined with increasing resolution, however the overall structure remains the same.

\begin{figure}
   \begin{center}
      \includegraphics[width=\columnwidth]{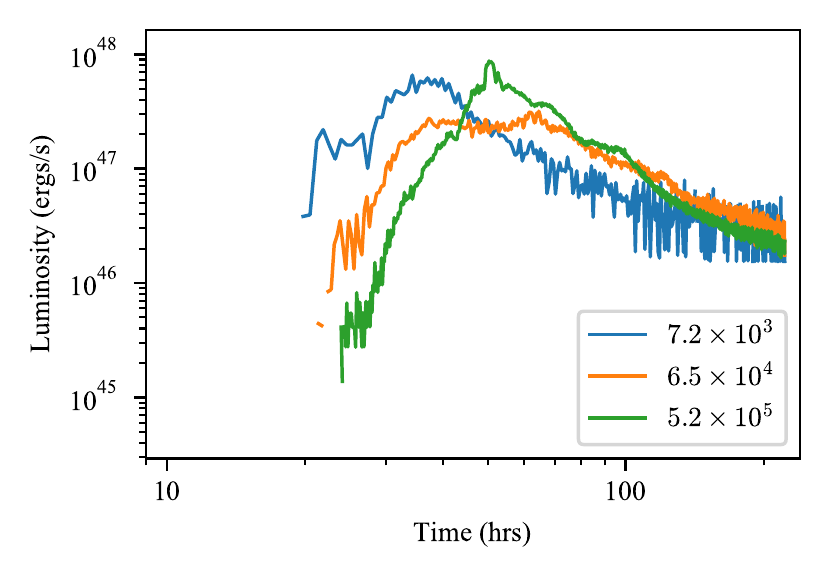}
      \caption{Comparison of $L_\mathrm{acc}$ at three different resolutions (number of particles) in Sim.~1. For lower resolutions, the accretion rate is higher during the early stage of the fallback and then lower at late times.}
      \label{fig:resolution-lacc}
   \end{center}
\end{figure}

Figure~\ref{fig:resolution-lacc} compares our approximate luminosity from the accretion rate at each resolution. At early times (see top row of Figure~\ref{fig:resolution}) differences are mainly caused by progressively better resolving of the circularisation process. The increased accretion luminosity at the highest resolution (green curve) seen at $t\approx 50$ hrs (see middle row of Figure~\ref{fig:resolution}) is mainly driven by an increased amount of material on plunging orbits. At late times, $t>100$ hrs, these differences become irrelevant (bottom row of Figure~\ref{fig:resolution}).


\bsp	
\label{lastpage}
\end{document}